\documentclass[10pt,twocolumn,showpacs,amsmath,amssymb,floatfix,superscriptaddress]{revtex4-2}
\usepackage{graphicx}
\usepackage{float}
\usepackage{dcolumn}
\usepackage{bm}
\usepackage{color}
\usepackage{txfonts}
\usepackage{microtype}
\usepackage{hyperref}
\usepackage{mathrsfs}
\usepackage{easyReview}
\usepackage{amsmath}
\usepackage{extarrows}
\usepackage{mathtools}
\usepackage{xcolor}
\usepackage{slashed}
\usepackage{amsmath}

\newcommand{\DWtag}{\mathrm{DW}}
\newcommand{\VDWtag}{\mathrm{V}\text{-}\mathrm{DW}}
\setreviewson 
\begin{document}

\title{Nuclear excitation via inelastic scattering of low-energy vortex electrons}

\author{Jia-Lin Zhang}
\affiliation{Ministry of Education Key Laboratory for Nonequilibrium Synthesis and Modulation of Condensed Matter, State key laboratory of electrical insulation and power equipment, Shaanxi Province Key Laboratory of Quantum Information and Quantum Optoelectronic Devices, School of Physics, Xi'an Jiaotong University, Xi'an 710049, China}
\author{Zhi-Wei Lu}
\affiliation{Ministry of Education Key Laboratory for Nonequilibrium Synthesis and Modulation of Condensed Matter, State key laboratory of electrical insulation and power equipment, Shaanxi Province Key Laboratory of Quantum Information and Quantum Optoelectronic Devices, School of Physics, Xi'an Jiaotong University, Xi'an 710049, China}
\author{Mamutjan Ababekri}
\affiliation{Ministry of Education Key Laboratory for Nonequilibrium Synthesis and Modulation of Condensed Matter, State key laboratory of electrical insulation and power equipment, Shaanxi Province Key Laboratory of Quantum Information and Quantum Optoelectronic Devices, School of Physics, Xi'an Jiaotong University, Xi'an 710049, China}
\author{Yuanbin Wu}
\email{yuanbin@nankai.edu.cn}
\affiliation{School of Physics, Nankai University, Tianjin 300071, China}
\author{Jian-Xing Li}
\email{jianxing@xjtu.edu.cn}
\affiliation{Ministry of Education Key Laboratory for Nonequilibrium Synthesis and Modulation of Condensed Matter, State key laboratory of electrical insulation and power equipment, Shaanxi Province Key Laboratory of Quantum Information and Quantum Optoelectronic Devices, School of Physics, Xi'an Jiaotong University, Xi'an 710049, China}
\affiliation{Department of Nuclear Physics, China Institute of Atomic Energy, P. O. Box 275(7), Beijing 102413, China}

\date{\today}

\begin{abstract}
Vortex particles carrying orbital angular momenta (OAMs) have found important applications in broad fields.
However, the experimental verification of OAM transfer at the nuclear scale remains a great challenge.
Here, we put  forward a novel method to probe such OAM transfer through nuclear excitation via inelastic scattering of low-energy vortex electrons.
We develop a Dirac distorted-wave Born approximation framework that incorporates the incident-electron OAM and a nonperturbative treatment of the Coulomb field, and apply it to $^{229}\mathrm{Th}$.
We find that the vortex and non-vortex electrons yield opposite angular distributions,  attributed to the OAM-modified selection rule and the Coulomb-induced redistribution of partial-wave strengths, providing an angle-resolved signature. 
Moreover, the vortex electron exhibits topological protection in the nuclear Coulomb field.
Our method offers a route to probing nuclear-scale OAM transfer and deepens our understanding of the topological properties of vortex particles.
\end{abstract}

\maketitle
Vortex particles, described by wave packets with helical phases and carrying quantized orbital angular momenta (OAMs) along the propagation axis~\cite{bliokh2017theory,lloyd2017electron,ivanov2022promises,allen1992orbital,knyazev2018beams}, have given rise to new phenomena in various fields, such as quantum information~\cite{fetter2001vortices,ivanov2012creation}, condensed matter~\cite{edstrom2016elastic,grillo2017observation}, optical physics~\cite{padgett2017orbital,hernandez2017generation}, atomic physics~\cite{vanBoxem2015inelastic,ivanov2023studying,han2023attosecond, Lange2022octupoleint, Babiker2018atom}, and nuclear physics~\cite{wu2022dynamical,lu2023manipulation,lu2025angular,kirschbaum2024photoexcitation,lu2026nuclear}. 
As an additional degree of freedom, such OAM can be transferred to atoms or nuclei through the modification of selection rules~\cite{vanBoxem2015inelastic,lu2023manipulation}, thereby opening new avenues for controlling quantum transitions.
Studies of atomic excitation have shown that absorption of photons with a nonzero OAM by atoms can excite multipole transitions that are otherwise suppressed for plane-wave photons~\cite{schulz2020generalized,duan2019selection,scholz2014absorption}. 
Recent work has experimentally confirmed that vortex photons can transfer their OAM to bound electrons~\cite{schmiegelow2016transfer}. 
In nuclear physics, theoretical work has predicted that vortex particles could selectively access  giant resonances (GRs) with higher multipolarities~\cite{lu2023manipulation,lu2025angular}, control isomer depletion in nuclear excitation by electron capture (NEEC)~\cite{wu2022dynamical}, and drive the dipole-forbidden $^{229}$Th nuclear clock transition~\cite{kirschbaum2024photoexcitation}.
However, OAM transfer at the nuclear scale, which underlies all these predicted effects, has not yet been experimentally demonstrated~\cite{ivanov2022promises}.

\begin{figure*}[!th]
\centering
\vspace{-0.3cm}
\includegraphics[trim=0 13 0 0,clip,width=\textwidth]{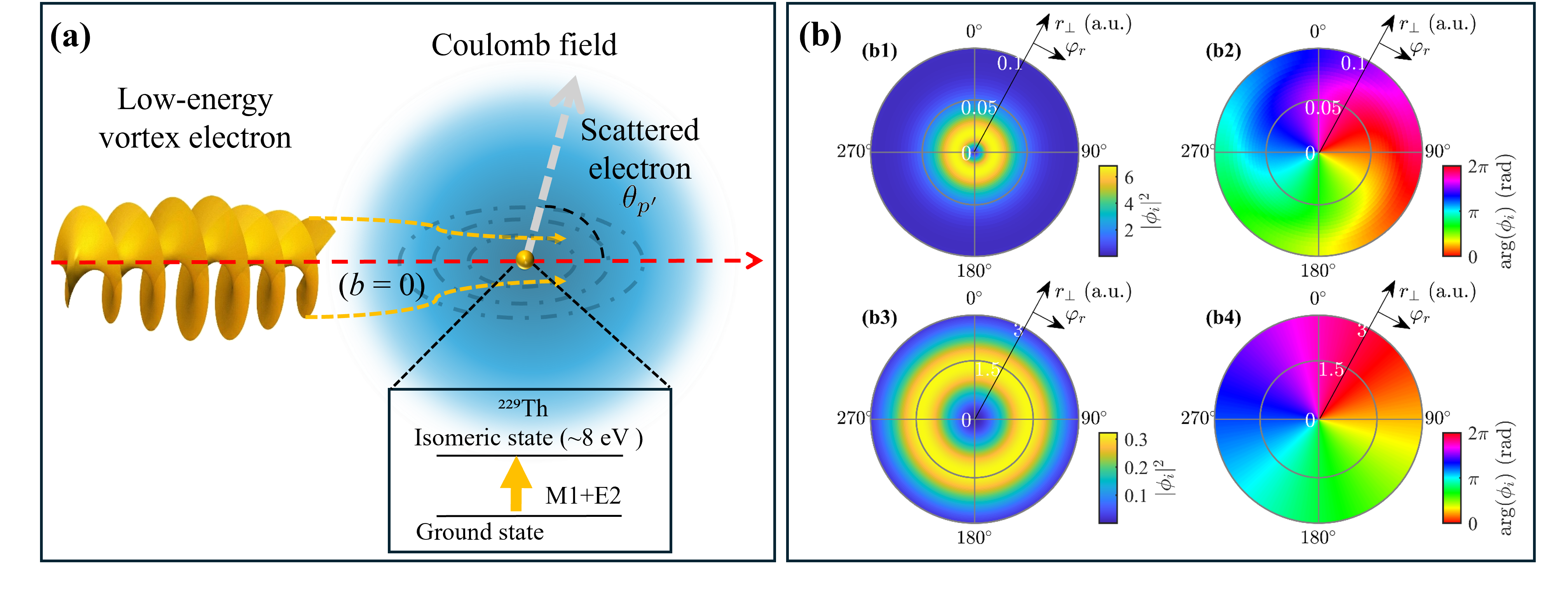}
\caption{(a) A low-energy vortex electron, incident on the $^{229}\mathrm{Th}$ nucleus along the beam axis, traverses the Coulomb field, excites the $\sim8\,\mathrm{eV}$ isomeric state through a mixed M1+E2 transition, and is scattered into the polar angle $\theta_{p'}$.
(b1), (b2) The transverse probability density $|\phi_i|^2$ and phase $\arg(\phi_i)$ of the incident vortex electron wave function in the V-DW case (Coulomb distortion). (b3), (b4) The corresponding quantities in the V-PW case (without Coulomb distortion). Parameters: TAM projection $m=3/2$,  helicity $\sigma_i=1/2$, incident energy $\varepsilon_i = 100~\mathrm{eV}$, opening angle $\theta_p = 30^\circ$, impact parameter $b = 0$.
}
\label{fig_overview}
\end{figure*}

Experimentally,  advances in wave-front-shaping techniques~\cite{uchida2010generation, verbeeck2010production,mcmorran2011electron,clark2015controlling} have enabled the generation of  vortex photons spanning eV to keV energies~\cite{maruyama2022generation,shen2019optical,peele2002observation,terhalle2011generation,gariepy2014creating,hemsing2013coherent} and vortex electrons with kinetic energies of hundreds of eV to hundreds of keV~\cite{vanacore2019ultrafast, beche2014magnetic,grillo2015holographic,rihacek2021beam}.
Notably, the recent experiment has further extended vortex $\gamma$-photon generation to the sub-MeV regime via inverse Compton scattering~\cite{wei2026experimental}.
However, the generation of vortex $\gamma$ photons in the GR energy range of approximately 10--30 MeV~\cite{harakeh2001giant,colo2023theoretical,zilges2022photonuclear} and vortex electrons at hundreds of MeV for extracting GR strengths remains a formidable challenge~\cite{ivanov2022promises,baturin2022evolution,sizykh2024transmission,ababekri2024generation,jentschura2011generation}.
In addition, as resonant processes, NEEC and photon-absorption require  incident energy to match the specific nuclear transition energy within its natural linewidth~\cite{palffy2006theory, beeks2021thorium}.
This condition is experimentally demanding even for conventional NEEC~\cite{chiara2018isomer, ding2026isomer,wu201993mmo,guo2022probing} and direct laser excitation of $^{229}\mathrm{Th}$~\cite{tiedau2024laser,elwell2024laser,zhang2024frequency}.
Therefore, based on these existing theoretical studies, the requirements of high-energy vortex particles or strict resonance energy matching limit the experimental demonstration of OAM transfer at the nuclear scale.

\begin{figure}[tb]
\centering
\includegraphics[trim=0 5 0 0,clip,width=\columnwidth]{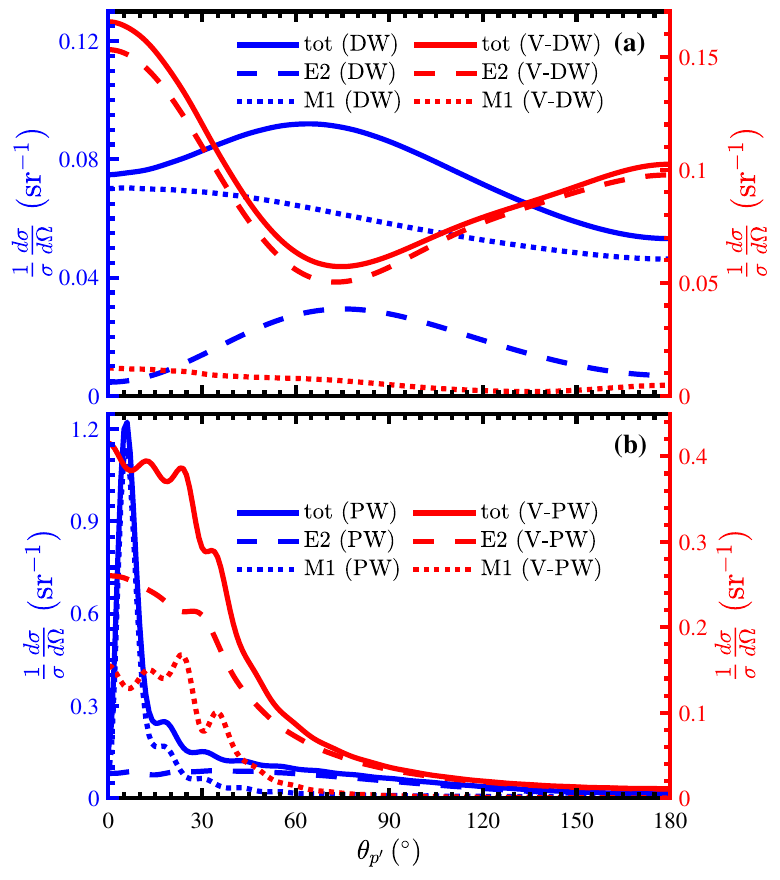}
\caption{The normalized differential cross section $(1/\sigma)d\sigma/d\Omega$ vs the scattered polar angle $\theta_{p'}$ for the M1, E2, and total (M1+E2) transitions, calculated in the (a) Dirac-DWBA and (b) Dirac-PWBA frameworks. 
Solid, dashed, and dotted curves correspond to the total, E2, and M1 transitions, respectively.
Blue curves denote the non-vortex cases, whereas red curves denote the vortex cases.
Parameters are the same as in Fig.~\ref{fig_overview}.
}
\label{fig_angdist}
\end{figure}

\begin{figure}[t]
    \centering
    \vspace{-0.3cm}
    \includegraphics[trim=0 8 0 0,clip,width=\columnwidth]{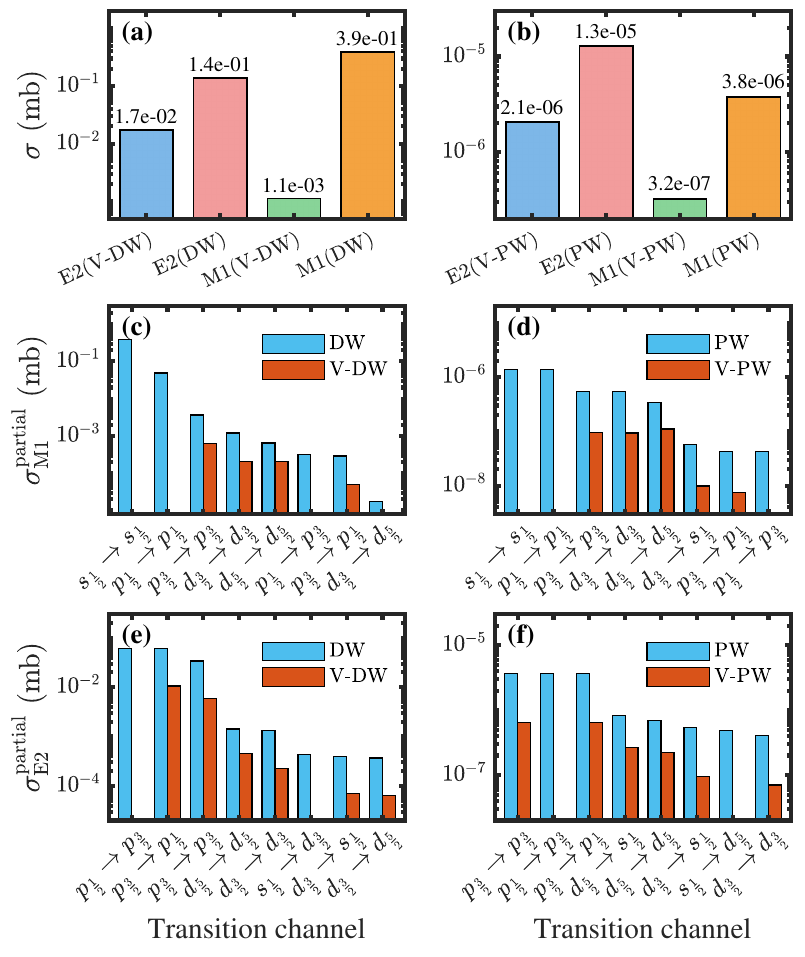}
    \caption{(a), (b) The cross sections for the E2 and M1 transitions within the Dirac-DWBA and Dirac-PWBA frameworks. (c), (d) The corresponding M1 partial-wave transition channels. (e), (f) The corresponding E2 partial-wave transition channels. Here, red (blue) bars represent the vortex (non-vortex) electron case. Parameters are the same as in Fig.~\ref{fig_overview}.
 }
\label{fig_filter}
\end{figure}

\begin{figure}[t]
\centering
\vspace{-0.3cm}
\includegraphics[trim=0 3 0 0,clip,width=\columnwidth]{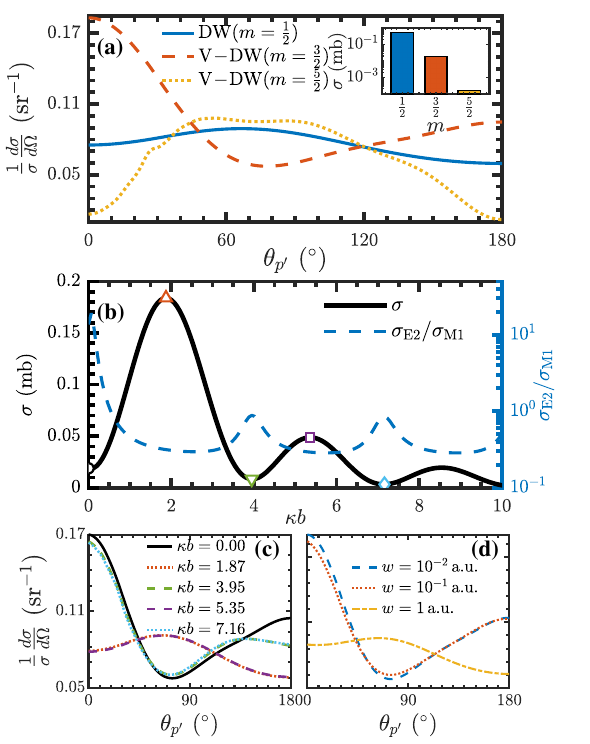}
\caption{Dependence of the angle-resolved signature on the beam and target parameters, calculated within the Dirac-DWBA framework. (a) The total (M1+E2) normalized differential cross section $(1/\sigma)\,d\sigma/d\Omega$ vs the scattered polar angle $\theta_{p'}$ under TAM projections $m=1/2$, $3/2$, and $5/2$ of the incident electrons. The inset shows their corresponding total cross sections $\sigma$. (b) The total cross section $\sigma$ and ratio $\sigma_{\mathrm{E2}}/\sigma_{\mathrm{M1}}$ as functions of the dimensionless impact parameter $\kappa b$, where $\kappa = p_i\sin\theta_p$. (c) The total normalized differential cross sections $(1/\sigma)\,d\sigma/d\Omega$  at $\kappa b=0$, 1.87, 3.95, 5.35, and 7.16. (d) The total normalized differential cross section $(1/\sigma)\,d\sigma/d\Omega$ for target width $w=10^{-2}$, $10^{-1}$ a.u. and 1 a.u. All other parameters are the same as in Fig.~\ref{fig_overview}.}
\label{fig:fig4}
\end{figure}

In this Letter, we put forward a novel method to probe such OAM transfer through nuclear excitation via nonresonant inelastic scattering of low-energy vortex electrons.
For low-energy electrons, the Coulomb field of the nucleus distorts the electron wave function, so that the standard plane-wave Born approximation fails~\cite{tkalya2020excitation,zhang2022nuclear,xu2024inelastic}.
To treat the nuclear Coulomb field nonperturbatively, we develop a  Dirac distorted-wave Born approximation (Dirac-DWBA) framework that incorporates the OAM of vortex electrons.
As a representative case, this framework is applied to the $^{229}$Th nucleus, whose low-lying ($\sim8\,\mathrm{eV}$) isomeric transition~\cite{seiferle2019energy,kroger1976features,thirolf2024thorium} is a magnetic dipole (M1) and electric quadrupole (E2) multipole mixture that is sensitive to angular-momentum selection rules.
The incident vortex electron propagates along the axis, traverses the nuclear Coulomb field, excites the nucleus and is scattered into a polar angle $\theta_{p'}$~[Fig.~\ref{fig_overview}(a)]. 
We find that the nuclear Coulomb field strongly reshapes the incident vortex electron wave function while preserving its phase singularity and OAM projection $m_\ell$. This topological protection ensures that the OAM is delivered through the Coulomb field to the nucleus~[Fig.~\ref{fig_overview}(b)]. 
More importantly, the angular distributions of scattered electrons induced by vortex and non-vortex electrons exhibit opposite trends, with the dominant multipole transition inverted from M1 to E2, providing an angle-resolved signature of OAM transfer at the nuclear scale~(Fig.~\ref{fig_angdist}).
This inversion is attributed to the modified selection rule and the Coulomb-induced redistribution of partial-wave strengths~(Fig.~\ref{fig_filter}).
Finally, we investigate how this signature depends on the beam and target parameters~(Fig.~\ref{fig:fig4}).
Our findings provide a promising route to probing nuclear-scale OAM transfer.
Throughout, atomic units are used ($\hbar = m_e = e =1$).

We describe the incident vortex electron as a Bessel wave packet specified by longitudinal momentum $p_z$, transverse momentum  $|\mathbf{p}_\perp| = \kappa$, helicity $\sigma_i$, and  total angular momentum (TAM) projection $m$ along the propagation axis~\cite{vanBoxem2015inelastic,ivanov2023studying}. 
In the non-relativistic limit, the OAM projection $m_\ell \simeq m - \sigma_i$~\cite{bliokh2011relativistic}.
In momentum space, the state is represented as a coherent superposition of plane-wave components configured on opening angle $\theta_p = \arctan(\kappa/p_z)$.
To treat the nuclear Coulomb field nonperturbatively, each plane-wave component is replaced by a Dirac distorted wave (DW) $|\phi_i\rangle_{\mathbf{p}_i}^{\sigma_i,\DWtag}(\mathbf{r})$ with asymptotic momentum $\mathbf{p}_i$, obtained by solving the Dirac equation in the Dirac-Hartree-Fock-Slater  potential $V_{\mathrm{DHFS}}(r) $ using the RADIAL package~\cite{salvat2019radial,liu2022isomeric}.
The resulting Dirac vortex distorted-wave (V-DW) state is~\cite{ivanov2015scattering,zaytsev2017radiative,zaytsev2018elastic,groshev2020brems} 
\begin{equation}
|\phi_i\rangle_{\kappa m p_z}^{\sigma_i,\VDWtag}(\mathbf{r}+\mathbf{b})=
\int \frac{d^{2}\mathbf{p}_\perp}{(2\pi)^2}\,
\alpha_{\kappa m}(\mathbf{p}_\perp)\, e^{i\mathbf{p}\cdot\mathbf{b}}\,
|\phi_i\rangle_{\mathbf{p}_i}^{\sigma_i,\DWtag}(\mathbf{r}),
\label{eq:vortex_state}
\end{equation}
where $\alpha_{\kappa m}(\mathbf{p}_\perp)$ is the vortex amplitude, and $\mathbf{b}=b(\cos\varphi_b,\sin\varphi_b)$ is the transverse displacement of the nucleus from the beam axis (details in~\cite{supplementalMaterial}).

Within the Dirac-DWBA framework, the differential cross section $\left.d\sigma/d\Omega\right|_{\sigma_i}^{\VDWtag}$ is obtained from the transition amplitude $\langle f|H_{\rm int}|i\rangle^{\VDWtag}$, which can be written as a coherent superposition of the DW transition amplitude $\langle f|H_{\rm int}|i\rangle^{\DWtag}$ (details in~\cite{supplementalMaterial}):
\begin{equation}
\begin{aligned}
\langle f | H _ { i n t } | i\rangle^{\VDWtag}=
\int \frac{d^{2}\mathbf{p}_\perp}{(2\pi)^2}\,
\alpha_{\kappa m}(\mathbf{p}_\perp)\, e^{i\mathbf{p}\cdot\mathbf{b}}\, \langle f | H _ { i n t } | i\rangle^{\DWtag}.
\label{eq_selection}
\end{aligned}
\end{equation}
Here, $|i\rangle$ and $|f\rangle$ denote the initial and final states of the electron--nucleus system.
The angular momentum transfer between the electron and the nucleus satisfies the selection rule $m_{j_i}  - m_{j_f} = M_f - M_i$, due to the conservation of angular momentum. Here, $m_{j_i, j_f}$ is the TAM projection of the incident or scattered electron partial wave that interacts with the nucleus, and $M_{i, f}$ denotes the magnetic quantum number of the initial or the final state of the nucleus.
In the DW case, the incident electron is expanded in partial waves with the helicity $\sigma_i$, so that the selection rule simplifies to $\sigma_i - m_{j_f} = M_f - M_i$.
The transition amplitude $\langle f|H_{\rm int}|i\rangle^{\VDWtag}$ involves the Bessel factor $J_{m-m_{j_i}}(\kappa b)$, which originates from the wave function of the incident vortex electron.
When the nucleus is on-axis (b=0), this Bessel factor reduces to $J_{m-m_{j_i}}(0)=\delta_{m,m_{j_i}}$, yielding $m_{j_i}=m$, which modifies the selection rule to $m - m_{j_f} = M_f - M_i$.
Since $|m_{j_i}|\leq j_i$, all incident electron partial waves with $j_i<|m|$ are forbidden.
For $b\neq0$, the transition amplitude $\langle f|H_{\rm int}|i\rangle^{\VDWtag}$ contains coherent contributions from all TAM projections $m_{j_i}=-j_i,-j_i+1,\ldots,j_i$ of each incident partial wave, weighted by the Bessel factor $J_{m-m_{j_i}}(\kappa b)$.
When the Coulomb field $V_{\rm DHFS}(r)$ is neglected, this theory returns to the conventional Dirac-PWBA framework, where PW and V-PW denote the cases of non-vortex and vortex electrons, respectively (details in~\cite{supplementalMaterial}).

\textit{Topological protection in the Coulomb field.---}
Figure 1(b) displays the transverse probability density $|\phi_i(\mathbf{r})|^2$ and phase $\arg[\phi_i(\mathbf{r})]$ of the vortex electron with and without the nuclear Coulomb field in the transverse plane of $^{229}$Th.
As shown in Figs.~\ref{fig_overview}(b1) and \ref{fig_overview}(b3), the probability density of the vortex electron exhibits a hollow distribution with azimuthal symmetry.
The nuclear Coulomb field strongly focuses the transverse profile of the electron around the nucleus, enhancing the peak probability density by more than an order of magnitude.
The corresponding phase distributions presented in Figs.~\ref{fig_overview}(b2) and \ref{fig_overview}(b4) exhibit a phase singularity at $b=0$, around which the phase accumulates by $2\pi$ over a closed loop.
Although the Coulomb potential strongly distorts the phase profile as the vortex electron traverses the nuclear Coulomb field, it preserves the phase singularity and  the OAM projection $m_\ell$ in the axially symmetric configuration, manifesting its topological protection. 
The preservation of vortex structure has previously been examined for electrons traversing macroscopic guiding fields such as magnetic lenses and solenoids~\cite{baturin2022evolution,sizykh2024transmission}. 
Here, the topological protection ensures that the OAM is reliably  delivered  through the Coulomb field to the nucleus, offering a potential probe of nuclear structure.

\textit{Angle-resolved signature of OAM transfer.---}
Figure~\ref{fig_angdist} shows the normalized differential cross section $(1/\sigma)\,d\sigma/d\Omega$  induced by non-vortex and vortex electrons, calculated within the Dirac-DWBA [Fig.~\ref{fig_angdist}(a)] and Dirac-PWBA [Fig.~\ref{fig_angdist}(b)] frameworks. 
The total angular distribution for the DW case first gradually increases and then decreases with increasing $\theta_{p'}$, peaking at $\theta_{p'}\simeq70^\circ$.
The M1 component dominates the distribution, while the E2 component exhibits a strong peak around  $\theta_{p'}\simeq70^\circ$, giving rise to the peak in the total angular distribution.
By contrast, the total angular distribution for the V-DW case  first drops to a minimum at $\theta_{p'}\simeq70^\circ$ and then rises with increasing $\theta_{p'}$, which is opposite to the trend in the DW case.
Here, the M1 component is suppressed  more strongly relative to the E2 component, so that the E2 component governs the total distribution.
In the Dirac-PWBA framework,  the total angular distribution  shows a narrow peak within $\theta_{p'}\lesssim 30^\circ$ for the PW case, where the M1 component dominates, while the E2 component governs larger angles.
In the V-PW case, the M1 component is also suppressed more strongly than the E2 component, so that the E2 component governs the total angular distribution, which decreases smoothly with $\theta_{p'}$.

The angular distributions induced by non-vortex and vortex electrons exhibit opposite trends over almost the entire range of scattering angles in the Dirac-DWBA framework, where the dominant multipole is inverted from M1 to E2, whereas their trends differ only for $\theta_{p'}\lesssim30^\circ$ in the Dirac-PWBA framework. 
This opposite trend provides a distinctive angle-resolved  signature of OAM transfer at the nuclear scale.
Moreover, the signature is most pronounced at large scattering angles,  where the vortex and non-vortex cases differ most strongly and the scattered electrons are naturally separated from the incident beam, facilitating identification of the OAM transfer by measuring the angular distribution of the scattered electrons in this regime.

\textit{Mechanism of the multipole inversion.---}To clarify the mechanism of the multipole inversion in Fig.~\ref{fig_angdist}, we compare the integrated E2 and M1 cross sections  within the Dirac-DWBA [Fig.~\ref{fig_filter}(a)] and Dirac-PWBA [Fig.~\ref{fig_filter}(b)] frameworks.
From the DW to the  V-DW case, the M1 transition is suppressed by a factor of $\simeq 340$ while the E2 transition decreases by only a factor of $\simeq 7.9$, owing to the nonzero OAM of vortex electrons.
Since the M1 transition is suppressed far more strongly than the E2 transition, the dominant multipole is inverted from M1 in the DW case to E2 in the V-DW case, with the ratio $\sigma(\mathrm{E2})/\sigma(\mathrm{M1})$ increasing from $0.36$ to $15.6$.
By contrast, the M1 and E2 transitions are suppressed by comparable factors of $\simeq 12$ and $\simeq 6.2$ from the PW to the V-PW case so that E2 remains dominant and $\sigma(\mathrm{E2})/\sigma(\mathrm{M1})$ merely grows from $3.40$ to $6.36$.
In addition, the cross sections within the Dirac-DWBA framework are about four orders of magnitude larger than those within the Dirac-PWBA framework.
The multipole inversion disappears in the Dirac-PWBA framework, indicating that it depends on the nuclear Coulomb field.

Figures~\ref{fig_filter}(c)--\ref{fig_filter}(f) show the corresponding partial-wave transition channels for the M1 and E2 transitions.
For the vortex electron ($b=0$, $m=3/2$), only partial waves with $m_{j_i}=3/2$ are allowed, and since $|m_{j_i}|\leq j_i$, all entrance channels with $j_i<3/2$ are forbidden.
In the DW case, the M1 transition is mainly concentrated in the $s_{1/2}\to s_{1/2}$ and $p_{1/2}\to p_{1/2}$ channels [Fig.~\ref{fig_filter}(c)].
These two $j_i=1/2$ entrance channels are forbidden in the V-DW case, which leads to a strong suppression of the M1 component.
By contrast, in the DW case, the E2 component is dominated by the $p_{1/2}\to p_{3/2}$, $p_{3/2}\to p_{1/2}$, and $p_{3/2}\to p_{3/2}$ channels [Fig.~\ref{fig_filter}(e)].
While in the V-DW case, only the $p_{1/2}\to p_{3/2}$ channel is forbidden, whereas the other two channels with $j_i=3/2$ remain allowed.
Therefore, the E2 transition is suppressed much less than the M1 transition, so that the dominant multipole is inverted from M1 to E2 between the DW and V-DW cases.
However, in the Dirac-PWBA framework, the M1 and E2 transitions in the PW case are distributed among different $j_i$ channels with similar contributions [Fig.~\ref{fig_filter}(d) and (f)], rather than being strongly concentrated in the $j_i=1/2$ channels as in the DW case. As a result, the suppression of the M1 and E2 transitions caused by forbidding the $j_i=1/2$ channels is comparable and far less than that in the Dirac-DWBA framework, so that E2 remains the dominant multipole in both PW and V-PW cases.
By enhancing the contribution of low-$j_i$ partial wave channels for the M1 strength, the nuclear Coulomb field inverts the dominant multipole from M1 to E2 between the vortex and non-vortex cases in the Dirac-DWBA framework.

\textit{Dependence on beam and target parameters.---}In Fig.~\ref{fig:fig4}, we investigate how the angle-resolved signature depends on the TAM projection, impact parameter, and target size.
Here, only one parameter is changed in each case.
Figure~\ref{fig:fig4}(a) shows the normalized angular distributions of the DW case and the V-DW cases with different TAM projections.
Relative to the DW case, the V-DW case with  $m=3/2$  exhibits an opposite trend, whereas that with $m=5/2$ shows a similar trend.
Moreover, the inset shows the cross section decreases sharply as the TAM projection increases, due to the forbidding of all entrance channels with $j_i<|m|$.
Therefore, the $m=3/2$ case provides a more favorable signature of OAM transfer at the nuclear scale.
Figure~\ref{fig:fig4}(b) shows the total cross section and the ratio $\sigma_{\mathrm{E2}}/\sigma_{\mathrm{M1}}$ as functions of the dimensionless impact parameter $\kappa b$.
The total cross section $\sigma$ oscillates  with increasing $\kappa b$, reaching maxima at $\kappa b =1.87$ and $5.35$, and minima at $\kappa b =0$, $3.95$ and $7.16$, while the ratio $\sigma_{\mathrm{E2}}/\sigma_{\mathrm{M1}}$ varies inversely with $\sigma$.
These oscillations stem from the Bessel weights $J_{m-m_{j_i}}(\kappa b)$ in Eq.~\eqref{eq_selection}, which control the contribution of the $j_i=1/2$ entrance channels that dominate the M1 transition.
At the minima ($\kappa b = 3.95$ and $7.16$), the M1 component is strongly suppressed and  $\sigma_{\mathrm{E2}}/\sigma_{\mathrm{M1}}$ reaches its maxima. 
Figure~\ref{fig:fig4}(c) shows the normalized angular distributions at the $\kappa b$ values corresponding to the maxima and minima of $\sigma$.
At the maxima, the angular distribution shows a trend similar to the DW case, whereas at the minima this trend is nearly reversed.
Therefore, for some cases with $b \neq 0$, the reversed  trend remains a distinctive signature of OAM transfer.
For a mesoscopic target with a Gaussian spatial distribution, Fig.~\ref{fig:fig4}(d) shows the normalized angular distributions at different target sizes $w$, averaged over the impact parameter $\mathbf{b}$.
The opposite trend is retained for $w=10^{-2}$ and $10^{-1}$ a.u., whereas the distribution approaches the DW case for $w=1$ a.u.
This is because the average over larger $w$ includes sizable finite-b contributions, where the M1-dominant $j_i=1/2$ channels are no longer excluded.
Thus, the opposite trend remains visible as long as the target is smaller than the transverse scale of the vortex beam.
Furthermore, the opposite trend still persists for opening angles ranging from $15^\circ$ to 
$60^\circ$ and incident energies from $10^2$ to $10^5$~eV (see Fig.~S1 in~\cite{supplementalMaterial}). 

\textit{Experimental feasibility.---}
We now turn to the experimental suggestions for this method.
The total cross section reaches the value $\sim1.8\times10^{-2}~{\rm mb}$ (the parameters in Fig.~\ref{fig_overview}).
In order to obtain a reaction rate per nucleus, we multiply this value by the vortex beam flux.
We assume here the generic flux of $10^{25}~{\rm cm}^{-2}{\rm s}^{-1}$~\cite{beche2017efficient,reimer2008transmission}, yielding the value of approximately $15~\mathrm{d}^{-1}$.
It should be noted that the definition of the incident flux for vortex particles, and hence the corresponding cross section, is not unique and can depend on the adopted flux prescription~\cite{ivanov2015scattering,ivanov2020kinematic}.
Here we adopt the most conservative prescription by normalizing to the beam-averaged flux~\cite{lu2025angular}, so that the cross section  should be regarded as a conservative estimate.
The experimental goal is to preserve the opposite angular trend.
For opening angles $\theta_p=30^\circ$, $15^\circ$, and even $5^\circ$, the required target widths are $5.3\times 10^{-2}$, $1.0\times 10^{-1}$, and $3.0\times 10^{-1}$~\AA, respectively, which should be accessible with current vortex-beam focusing techniques~\cite{bliokh2017theory,beche2017efficient}.
In addition, the energy loss of $\sim8\,\mathrm{eV}$ can separate the nuclear-excitation signal from elastic scattering. 
The main challenge is the atomic electronic background in the same energy-loss region, which can be suppressed by detecting the scattered electron in coincidence with the subsequent nuclear de-excitation.

In conclusion, applying our method to $^{229}$Th, we find that vortex and non-vortex electrons induce opposite angular distributions, with the dominant multipole inverted from M1 to E2, providing an angle-resolved signature of nuclear-scale OAM transfer that remains robust over a suitable range of beam and target parameters.
Moreover, the vortex electron preserves its phase singularity and OAM projection while traversing the nuclear Coulomb field. 
Our method thus offers a route to probing nuclear-scale OAM transfer and deepens our understanding of the topological properties of vortex particles.

{\it Acknowledgments---}The work is supported by the National Natural Science Foundation of China (Grants No. 12425510, No. U2267204, No. 12441506, No. 12505276, No. 12475122), the National Key Research and Development (R\&D) Program (Grant No. 2024YFA1610900), the Science Challenge Project (No. TZ2025012), and the Innovative Scientific Program of CNNC.

\bibliography{mybib}

\end{document}